\documentclass[aps,prl,twocolumn,showpacs,groupedaddress]{revtex4-1}

\usepackage{graphics}

\usepackage{epsfig}
\usepackage{amsmath}
\usepackage{amssymb}
\usepackage{amsfonts}
\usepackage{ dsfont }

\usepackage[debug]{mciteplus}

\usepackage{color}
\usepackage{ulem}

\newcommand{\be}{\begin{equation}}
\newcommand{\ee}{\end{equation}}

\begin{document}


\hsize\textwidth\columnwidth\hsize\csname@twocolumnfalse\endcsname

\title{Second harmonic coherent driving of a spin qubit in a Si/SiGe quantum dot}
\author{P. Scarlino$^1$, E. Kawakami$^1$, D. R. Ward$^2$, D. E. Savage$^2$, M. G. Lagally$^2$, Mark Friesen$^2$, S. N. Coppersmith$^2$, M. A. Eriksson$^2$ and L. M. K. Vandersypen$^1$}
\affiliation{
$^1$Kavli Institute of Nanoscience, TU Delft, Lorentzweg 1, 2628 CJ Delft, The Netherlands\\
$^2$University of Wisconsin-Madison, Madison, WI 53706, USA\\
}
\date{\today}
\vskip1.5truecm
\begin{abstract} We demonstrate coherent driving of a single electron spin using second harmonic excitation in a Si/SiGe quantum dot. Our estimates suggest that the anharmonic dot confining potential combined with a gradient in the transverse magnetic field dominates the second harmonic response. As expected, the Rabi frequency depends quadratically on the driving amplitude and the periodicity with respect to the phase of the drive is twice that of the fundamental harmonic. The maximum Rabi frequency observed for the second harmonic is just a factor of two lower than that achieved for the first harmonic when driving at the same power. Combined with the lower demands on microwave circuitry when operating at half the qubit frequency, these observations indicate that second harmonic driving can be a useful technique for future quantum computation architectures. \end{abstract} 
\textcolor{red}{} 
\maketitle 
Controlled two-level quantum systems are essential elements for quantum information processing. A natural and archetypical controlled two-level system is the electron spin doublet in the presence of an external static magnetic field~\cite{Loss1998,Hanson2007}. The common method for driving transitions between the two spin states is magnetic resonance, whereby an a.c.\ magnetic field ($B_{a.c.}$) is applied transverse to the static magnetic field ($B_{ext}$), with a frequency, $f_{MW}$, matching the spin Larmor precession frequency $f_{L}=g\mu_BB_{tot}/h$ ($h$ is Planck's constant, $\mu_B$ is the Bohr magneton and $B_{tot}$ the total magnetic field acting on the spin). When the driving rate is sufficiently strong compared to the dephasing rates, coherent Rabi oscillations between the ground and excited state are observed. 

Both spin transitions and Rabi oscillations can be driven not just at the fundamental harmonic but also at higher harmonics; i.e., where the frequency of the transverse a.c. field is an integer fraction of the Larmor frequency, $f_{MW} = f_L/n$, with $n$ an integer. Second or higher harmonic generation involves non-linear phenomena. Such processes are well known and explored in quantum optics using non-linear crystals \cite{Franken1961} and their selectivity for specific transitions is exploited in spectroscopy and microscopy \cite{Heinz1982,Shen1989,Denk1990, Xu1996,Konig2000}. Two-photon transitions have been extensively explored also in superconducting qubit systems \cite{Nakamura2001, Wallraff2003, Oliver2005,Shevchenko2012}. In cavity QED systems, a two-photon process has the advantage that it allows the direct transition from the ground state to the second excited state, which is forbidden in the dipole transition by the selection rules \cite{Poletto2012}. 

For electron spin qubits, it has been predicted that the non-linear dependence of the $g$-tensor on applied electric fields should allow electric-dipole spin resonance (EDSR) at subharmonics of the Larmor frequency for hydrogenic donors in a semiconductor~\cite{De2009,Pingenot2010}. For electrically driven spin qubits confined in a (double) quantum dot, higher-harmonic driving has been proposed that takes advantage of an anharmonic dot confining potential \cite{Rashba2011,Nowak2012, Osika2014,Danon2014} or a spatially inhomogeneous magnetic field~\cite{Szechenyi2014}. Although higher harmonics have been used in continuous wave (CW) spectroscopy for quantum dots hosted in GaAs, InAs, InSb and carbon nanotubes \cite{Laird2009,Pei2012,Laird2013,Stehlik2014,Nadj-Perge2012,Forster2015}, coherent spin manipulation using higher harmonics has not been demonstrated previously. 

In this letter we present experimental evidence of coherent second harmonic manipulation of an electron spin confined in a single quantum dot (QD) hosted in Si/SiGe quantum well. We show that this second-harmonic driving can be used for universal spin control, and we use it to measure the free-induction and Hahn-echo decay of the electron spin. Furthermore, we study how the second harmonic response varies with the microwave amplitude and phase, and comment on the nature of the non-linearity that mediates the second harmonic driving process in this system.  
\begin{figure} \includegraphics[width=8.5cm] {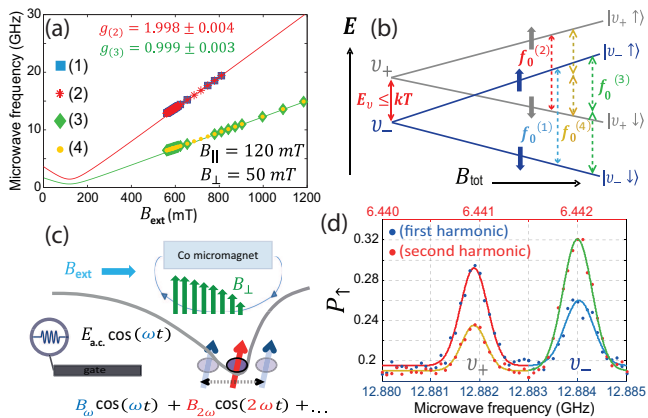} 
\caption{\label{fig:fig1} (a) Measured resonance frequencies as a function of externally applied magnetic field $B_{ext}$. The long microwave burst time $t_p =$700 $\mu$s $\gg T^\ast_2$ means that the applied excitation is effectively continuous wave (CW). The microwave source output power  was $P = - 33$ dBm to $-10$ dBm ($-20$ dBm to $-5$ dBm) for the case of fundamental (second) harmonic excitation, decreasing for lower microwave frequency in order to avoid power broadening. The red and green lines represent fits with the relation $hf=g\mu_B\sqrt{(B_{ext}-B_{||})^2+B_\bot^2}$ respectively to the resonance data labeled (2) and (3) (we excluded points with $B_{ext}<700$ mT from the fit because the micromagnet apparently begins to demagnetize there)~\cite{Kawakami2014}. (b) Schematic of the energy levels involved in the excitation process, as a function of the total magnetic field at the electron location. The dashed arrows correspond to the four transitions in panel (a), using the same color code. (c) Schematic of an anharmonic confinement potential, leading to higher harmonics in the electron oscillatory motion in response to a sinusoidally varying excitation. (d) Measured spin-up probability, $P_{\uparrow}$, as a function of applied microwave frequency, $f_{MW}$, for $B_{ext}$= 560.783 mT ($P = -30$ dBm for the fundamental response, $P = -12$ dBm for the second harmonics), averaged over 150 repetitions per point times 80 repeated frequency sweeps (160 mins in total). The frequency axis (in red on top) has been stretched by a factor of two for the second harmonic spin response (red datapoints). From the linewidths, we extract a lower bound for the dephasing time $T_2^{\ast(1)}$ = 760 $\pm$ 100 ns, $T_2^{\ast(2)}$ = 810 $\pm$ 50 ns, $T_2^{\ast(3)}$ = 750 $\pm$ 40 ns and $T_2^{\ast(4)}$ = 910 $\pm$ 80 ns. The Gaussian fits through the four peaks use the same color code as in panels (a) and (b).} 
\end{figure}  

The quantum dot is electrostatically induced in an undoped Si/SiGe quantum well structure, through a combination of accumulation and depletion gates (see Sec.~I of \cite{suppl.} for full details). The sample and the settings are the same as those used in Ref.~\cite{Kawakami2014}. A cobalt micromagnet next to the QD creates a local magnetic field gradient, enabling spin transitions to be driven by electric fields \cite{Obata2010,Kawakami2014}.

All measurements shown here are performed using single-shot readout via a QD charge sensor~\cite{Elzerman2004}. They make use of four-stage gate voltage pulses implementing (1) initialization to spin-down, (2) spin manipulation through all-electrical microwave excitation, (3) single-shot spin readout, and (4) a compensation/empty stage~\cite{Kawakami2014}. The results of many single-shot cycles are used to determine the spin-up probability, $P_{\uparrow}$, at the end of the manipulation stage. 

First we apply long, low-power microwave excitation to perform quasi-CW spectroscopy. Fig.~\ref{fig:fig1}(a) shows four observed spin resonance frequencies, $f_0^{(1)}$ through $f_0^{(4)}$, as a function of the external magnetic field. The resonances labeled (1) and (2) represent the response at the fundamental frequency. As in~\cite{Kawakami2014}, these two closely spaced resonances correspond to the electron occupying either of the two lowest valley states, both of which are thermally populated here. The other two resonances occur at exactly half the frequency of the first two, $f_0^{(1)}=2f_0^{(3)}$, $f_0^{(2)}=2f_0^{(4)}$, and represent the second harmonic response.

The effective $g$-factors extracted from the slopes for the second harmonic response are half those for the first harmonic response [see Fig.~\ref{fig:fig1}(a) inset]. The relevant energy levels and transitions as a function of the total magnetic field, $B_{tot}$, are visualized in Fig.~\ref{fig:fig1}(b), where the color scheme used for the resonances is the same as in Fig.~\ref{fig:fig1}(a). We see two sets of Zeeman split doublets, separated by the splitting between the two lowest-energy valleys, $E_v$. The transition between the Zeeman sublevels within each doublet can be driven by absorbing a single photon or two photons, as indicated by the single and double arrows. 

To drive a transition using the second harmonic, a non-linearity is required. In principle, several mechanisms can introduce such a non-linearity in this system (see Sec.~II of \cite{suppl.}). First, as schematically shown in Fig.~\ref{fig:fig1}(c), if the confining potential is anharmonic, an oscillating electric field of amplitude $E_{a.c.}$ and angular frequency $\omega=2\pi f_{MW}$ induces effective displacements of the electron wavefunction with spectral components at angular frequencies $n \omega$, with $n$ an integer. In analogy with non-linear optical elements, we can look at this process as generated by an effective non-linear susceptibility of the electron bounded to the anharmonic QD confinement potential.

The gradient in the transverse magnetic field in the dot region ($B_\bot$ in green) converts the electron motion into an oscillating transverse magnetic field of the form
\begin{equation} 
B^{a.c.}_\perp(t)=B_\omega \cos (\omega t)+B_{2\omega} \cos (2\omega t)+ \ldots \, 
\label{eq:Bperp} 
\end{equation}  
that can drive the electron spin for $\hbar\omega=E_z$, $2\hbar\omega=E_z$ and so forth~\cite{Rashba2011}.
A possible source of anharmonicity arises from the nonlinear dependence of the dipole moment between the valley (or valley-orbit) ground ($\upsilon_-$) and excited states ($\upsilon_+$) \cite{Gamble2013}, as a function of $E_{a.c.}$.

A second possible source of nonlinearity is a variation of the transverse field gradient, $\frac{dB_\perp}{dx,dy}$, with position [see Fig.~\ref{fig:fig1}(c)]. Even if the confining potential were harmonic, this would still lead to an effective transverse magnetic field containing higher harmonics, of the same form as Eq.\ref{eq:Bperp}.

A third possibility is that not only the transverse magnetic field but also the longitudinal magnetic field varies with position. This leads to an a.c. magnetic field which is not strictly perpendicular to the static field, which is in itself sufficient to allow second harmonic driving \cite{ Boscaino1971,Gromov2000}, even when the confining potential is harmonic and the field gradients are constant over the entire range of the electron motion.

However, simple estimates indicate that the second and third mechanisms are not sufficiently strong in the present sample to allow the coherent spin manipulation we report below (see Sec.~II of \cite{suppl.}).
We propose that the first mechanism is dominant in this sample, supported by our observation that the strength of the second harmonic response is sensitive to the gate voltages defining the dot.
\begin{figure} 
\includegraphics[width=8.5cm] {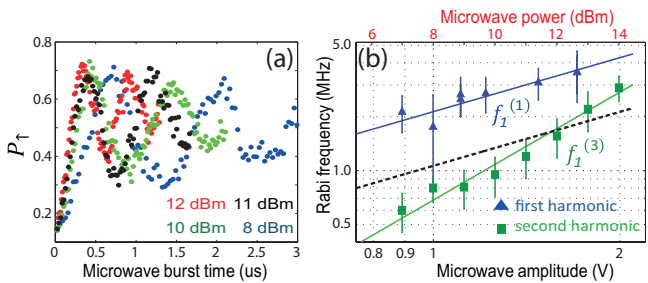} 
\caption{\label{fig:fig2} Rabi oscillations. (a) Measured spin-up probability, $P_\uparrow$, as a function of microwave burst time ($B_{ext} = 560.783$ mT, $f_{MW}$ = 6.4455 GHz) at four different microwave powers, corresponding to a rms voltage at the source of 998.8 mV, 1257.4 mV, 1410.9 mV, 1583.0 mV. (b) Rabi frequencies recorded at the fundamental harmonic, $f_0^{(1)}$ (blue triangles, adapted from \cite{Kawakami2014}), and at the second harmonic, $f_0^{(3)}$ (green squares), as a function of the microwave amplitude emitted from the source (top axis shows the corresponding power). For the second harmonic, the amplitude shown corresponds to a 5 dB higher power than the actual output power, to compensate for the 5 dB lower attenuation of the transmission line at 6 GHz versus 12 GHz (estimated by measuring the coax transmission at room temperature). The green solid (dashed black) line is a fit of the second harmonic data with the relation $\log(f^R) \propto 2\log(E_{a.c.})$ [$\log(f^R) \propto \log(E_{a.c.})$]. The large error bars in the FFT of the data in Fig.~\ref{fig:fig2}(a) arise because we perform the FFT on only a few oscillations. $B_{ext}$= 560.783 mT.} 
\end{figure} 
  
In Fig.~\ref{fig:fig1}(d) we zoom in on the four CW spin resonance peaks, recorded at low enough power to avoid power broadening (see Sec.~I of \cite{suppl.}). Fitting those resonances with Gaussians, we extract the dephasing times $T_2^{\ast,(1,2)}=\frac{\sqrt{2}\hbar}{\pi\delta f_{FWHM}^{(1,2)}}$, $T_2^{\ast,(3,4)}=\frac{\sqrt{2}\hbar}{2\pi\delta f_{FWHM}^{(3,4)}}$, giving values in the range of $750$ to $910$ns for all four resonances [see caption of Fig.~\ref{fig:fig1}(d)]. This directly shows that the linewidth (FWHM) extracted for the two-photon process is half that for the one-photon process, as expected \cite{Gromov2000,De2009,Szechenyi2014}.

From the relative peak heights in Fig.~\ref{fig:fig1}(d), we can estimate the ratio of the Rabi frequencies between the two peaks in each pair (see Sec.~I of \cite{suppl.}). In \cite{Kawakami2014}, we found that the relative thermal populations of the two valleys ($\epsilon^{(4)}/\epsilon^{(3)}$) were about $0.3\pm0.1$ to $0.7\pm0.1$. Given this, the ratio between the Rabi frequencies, $f_1$, extracted from the peak heights is $r_R(2ph)=f_1^{(4)}/f_1^{(3)} = 0.9\pm 0.2$ for the second harmonics. This is different from the ratio observed in \cite{Kawakami2014} for the fundamental harmonic, $r_R(1ph)=f_1^{(2)}/f_1^{(1)}=1.70\pm 0.05$ \cite{Note1}.

Such a difference is to be expected. The ratio $r_R(2ph)$ is affected by how the degree of anharmonicity in the confining potential differs between the two valleys. In contrast, $r_R(1ph)$ depends on how the electrical susceptibility differs between the two valleys \cite{Rahman2009}. In addition, since the valleys have different charge distributions~\cite{Gamble2013}, the microwave electric field couples differently to the two valley states, and this difference can be frequency dependent~\cite{Oosterkamp1997,Kawakami2013}. Because the second harmonic Rabi oscillations are driven at half the frequency of the Rabi oscillations driven at the fundamental, this frequency dependence also contributes to a difference between $r_R(1ph)$ and $r_R(2ph)$.
We note that the difference in Rabi frequency ratio between the 1-photon and 2-photon case demonstrates that the second harmonic response is not just the result of a classical up-conversion of the microwave frequency taking place before the microwave radiation impinges on the dot, but takes place at the dot itself. 
\begin{figure} 
\includegraphics[width=8.5cm] {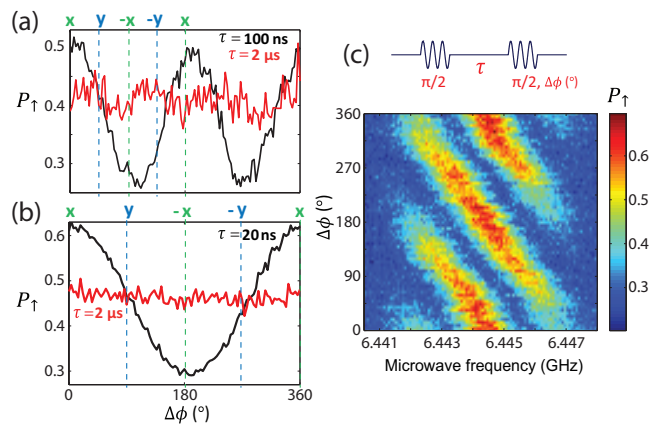} 
\caption{\label{fig:fig3} Phase control of oscillations. (a) Probability $P_\uparrow$ measured after applying two $\pi/2$ rotations via second harmonic excitation, as a function of the relative phase between the two microwave bursts, $\Delta \phi$. The two rotations are separated by $\tau = 100$ ns (black) and $\tau =$ 2 $\mu$s (red). ($P = 16.0$ dBm, $B_{ext} = 560.783$ mT, $f_{MW} = f_0^{(3)} = 6.44289$ GHz). (b) Similar to panel (a), but now driving the fundamental harmonic for $\tau = 20$ ns (black) and $\tau =$ 2 $\mu$s (red). ($P = 12.0$ dBm, $B_{ext} = 560.783$ mT, $f_{MW} = f_0^{(2)} = 12.88577$ GHz). Inset: Microwave pulse scheme used for this measurement. (c) Measured spin-up probability, $P_\uparrow$ (1000 repetitions for each point), as a function of $f_{MW}$ and the relative phase $\Delta\phi$ between two $\pi/2$ microwave bursts (130 ns, $P = 16.0$ dBm) for second harmonic excitation, with $\tau = 50$ ns. The measurement extends over more than 15 hours.} 
\end{figure}    

The second harmonic response also permits coherent driving, for which a characteristic power dependence is expected~\cite{Szechenyi2014,Gromov2000,Strauch2007}. Fig.~\ref{fig:fig2}(a) shows Rabi oscillations, where the microwave burst time is varied keeping $f_{MW}=f_0^{(3)}$ for different microwave powers. We note that the contribution to the measured spin-up oscillations coming from the other resonance, $(4)$, is negligible because the respective spin Larmor frequencies are off-resonance by 2 MHz, $f_1^{(3)}/f_1^{(4)}\approx 1$ and its population is $\sim$ three times smaller.

To analyze the dependence of the Rabi frequency on microwave power, we perform a FFT of various sets of Rabi oscillations similar to those in Fig.~\ref{fig:fig2}(a). Fig.~\ref{fig:fig2}(b) shows the Rabi frequency thus obtained versus microwave power for driving both at the second harmonic (green) and at the fundamental (blue), taken for identical dot settings \cite{Kawakami2014}. We see that for driving at the frequency of the second harmonic, the Rabi frequency is quadratic in the applied electric field amplitude (linear in power), as expected from theory~\cite{Szechenyi2014,Gromov2000,Strauch2007}. When driving at the fundamental resonance, the Rabi frequency is linear in the driving amplitude, as usual.
It is worth noting that at the highest power used in this experiment the Rabi frequency obtained from driving the fundamental valley-orbit ground state spin resonance is just a factor of two higher than the one from driving at the second harmonic. This ratio indicates that the use of second harmonic driving is quite efficient in our device.

A further peculiarity in coherent driving using second harmonics is seen when we vary the phase of two consecutive microwave bursts. Fig.~\ref{fig:fig3}(a) shows the spin-up probability following two $\pi/2$ microwave bursts with relative phase $\Delta\phi$, resonant with $f_0^{(3)}$ and separated by a fixed waiting time $\tau$. For short $\tau$, the signal oscillates sinusoidally in $\Delta\phi$ with a period that is half that for the single-photon case [compare the black traces in Figs.~\ref{fig:fig3}(a,b)].

Therefore, in order to rotate the electron spin around an axis in the rotating frame rotated by 90 degrees with respect to a prior spin rotation axis (e.g. a Y rotation following an X rotation in the rotating frame), we need to set $\Delta \phi$ to 45 degrees, instead of 90 degrees, when driving via the second harmonic. Of course, for $\tau \gg T_2^\ast$, the contrast has vanished, indicating that all phase information is lost during the waiting time [Fig.~\ref{fig:fig3}(a,b) red traces].  
Fig.~\ref{fig:fig3}(c) shows two-pulse measurements as in Fig.~\ref{fig:fig3}(a) as a function of frequency detuning and phase difference, where we can appreciate the extraordinary stability of the undoped device. 
\begin{figure} 
\includegraphics[width=8.5cm] {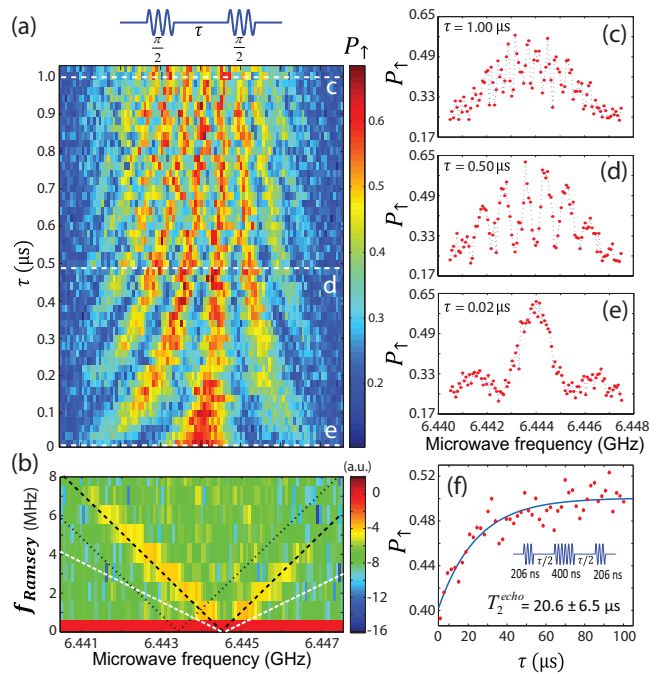} 
\caption{\label{fig:fig4} Ramsey fringes. (a) Measured spin-up probability, $P_\uparrow$, as a function of $f_{MW}$ and waiting time $\tau$ ($B_{ext} = 560.783$ mT, $P = 13.0$ dBm) between two $\pi/2$ pulses (130 ns) with equal phase, showing Ramsey interference. Each data point is averaged over 300 cycles. Inset: Microwave pulse scheme used for this measurement. (b) Fourier transform over the waiting time, $\tau$, of the data in panel (a), showing a linear dependence on the microwave frequency, with vertex at $f_{MW}=f_0^{(3)}$ and slope $f_{Ramsey}=2 \Delta f_{MW}$(black dashed lines). The expected position of the FFT of the signal arising from resonance $f_0^{(4)}$ is indicated by the dotted black line. For comparison, the white dashed line represents the relation $f_{Ramsey}=\Delta f_{MW}$. (c-d-e) Sections of the Ramsey interference pattern in (a) along the three white dashed lines; the respective waiting times are indicated also in the inset of each panel. (f) Measured spin-up probability as a function of the total free evolution time, $\tau$, in a Hahn echo experiment (pulse scheme shown in inset). The decay curve is fit well to a single exponential (blue). Here, $f_{MW}=f_0^{(3)}$, $B_{ext}$= 560.783 mT.} 
\end{figure}   

To probe further the coherence properties of the spin, we perform a free induction (Ramsey) decay measurement, see Fig.~\ref{fig:fig4}(a), as a function of frequency detuning and delay time, $\tau$, between the two bursts. The absence of a central frequency symmetry axis is due to the presence of two superimposed oscillating patterns, originating from the resonances at $f_0^{(3)}$ and $f_0^{(4)}$.
Figs.~\ref{fig:fig4}(c-e) show sections of the Ramsey measurement in Fig.~\ref{fig:fig4}(a), corresponding to different waiting times $\tau$ (see the white dashed lines). The visibility of the Ramsey fringes clearly decreases for longer waiting times between the two $\pi/2$ pulses. Fitting the decay of the visibility of the fringes as a function of $\tau$ with a Gaussian ($\propto\exp[-(t/T_2^\ast) ^2]$, see Sec.~I of \cite{suppl.}) we find $T_2^\ast = 780 \pm 110$ ns, in agreement with the value extracted from the linewidth.

Furthermore, and analogously to the observations of Fig.~\ref{fig:fig3}(a), we report a doubling effect in the frequency of the Ramsey oscillations, $f_{Ramsey}$, as a function of the detuning $\Delta f_{MW}=f_{MW}-f_0^{(3)}$.
Fig.~\ref{fig:fig4}(b) shows $f_{Ramsey}(\Delta f_{MW})$, extracted from the data in Fig.~\ref{fig:fig4}(a) via a FFT over the waiting time $\tau$. The black dashed line indicates the condition $f_{Ramsey}=2 \Delta f_{MW}$, closely overlapping with the position of the yellow peaks in the FFT. The black dotted line indicates the condition $f_{Ramsey}= 2(f_{MW}-f_0^{(4)})$; this second resonance is not very visible in the data, due to the lower population of the corresponding valley. For comparison, the white dashed line indicates the condition $f_{Ramsey}=\Delta f_{MW}$, which is the expected response when driving at the fundamental. 

Finally, we perform a spin echo experiment via second harmonic driving. Fig.~\ref{fig:fig4}(f) shows $P_\uparrow$ as a function of the total free evolution time $\tau$, for a typical Hahn echo pulse sequence (illustrated in the inset) consisting of $\pi/2$, $\pi$ and $\pi/2$ pulses applied along the same axis, separated by waiting times $\tau/2$ \cite{Hahn1950}. A fit to a single exponential yields $T_2^{echo}= 20.6\pm6.5 \mu$s at $B_{ext}$= 560.783 mT, compatible with the $T_2^{echo}$ of 23.0 $\pm1.2$ $\mu$s we observed at the same magnetic field when driving via the fundamental harmonic (see Sec.~I of \cite{suppl.}).   

To summarize, we report coherent second harmonic driving of an electron spin qubit defined in a Si/SiGe quantum dot, including universal single-spin rotations. The non-linearity that permits second harmonic driving is likely related to the anharmonic confining potential for the electron. This means that routine use of second harmonics for spin control would be possible provided there is sufficient control over the degree of anharmonicity of the confining potential. This could be very useful since driving a spin qubit at half its Larmor frequency would substantially simplify the microwave engineering required for high fidelity qubit control.   

\begin{acknowledgments} We acknowledge M. Rudner, A. P\'{a}lyi and T. Jullien for useful discussions, R. Schouten and M. J. Tiggelman for technical support. Research was supported by the Army Research Office (W911NF-12-0607), the European Research Council and the Dutch Foundation for Fundamental Research on Matter. E.K. was supported by a fellowship from the Nakajima Foundation.
\end{acknowledgments}

\appendix

\title{Supplementary material for ''Second Harmonic Coherent ...''}

\maketitle

\section{I. Device details and additional data} 

\subsection{Device and measurement technique}

\begin{figure}
\includegraphics[width=8.5cm] {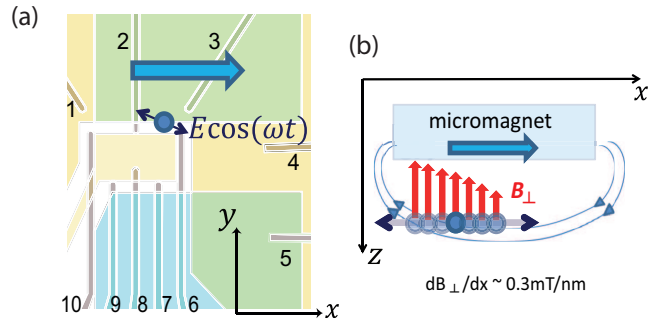}
\caption{\label{fig:figS1} (a) False-color device image showing a fabricated pattern of split gates, labeled 1-10. For this experiment we create a single quantum dot (estimated location indicated by a blue circle) and a sensing dot (gates 4 and 5). The current through the charge sensor is recorded in real time for a fixed voltage bias of 500 $\mu$eV. The voltage pulses and microwave excitation are applied to gate 2 and 6 respectively. Green semitransparent rectangles show the position of two 200-nm-thick Co micromagnets. The yellow-shaded areas show the location of two accumulation gates, one for the reservoirs and the other for the double quantum dot region. (b) Schematic side view of the stray magnetic field generated by a single micromagnet, completely magnetized along the x-axis; $B_\perp$ represents the component of the stray magnetic field perpendicular to the static field. The external magnetic field $B_{ext}$ is applied along the x-axis [blue arrow in both (a) and (b) panels].}
\end{figure}

The device used for all the measurements shown in the main text, is based on a 12 nm wide Si quantum well 37 nm below the surface in an undoped Si/SiGe heterostructure, with two layers of electrostatic gates [see Fig.~\ref{fig:figS1}(a)]. Two accumulation gates (in yellow) are used to induce a two-dimensional electron gas (2DEG) and a set of depletion gates (in gray), labeled 1-10 in Fig.~\ref{fig:figS1}(a), is used to form a single quantum dot in the 2DEG (on the right side) with a charge sensor next to it (made by gates 4, 5 and 6). Two 1 $\mu$m-wide, 200 nm-thick, and 1.5 $\mu$m-long Co magnets (in green), placed on top of accumulation gates (separated by an Al$_2$O$_3$ layer), provide a stray magnetic field [see Fig.~\ref{fig:figS1}(b)]. The sample is thermally anchored to the mixing chamber of a dilution refrigerator with base temperature 25 mK and the electron temperature estimated from transport measurements is 150 mK.

Microwave excitation, applied to gate 6 in this experiment, generates an a.c. electric field, $E_{a.c.}$, which makes the electron oscillate back and forth in the dot. Due to the gradient in the transverse magnetic field, $dB_{\bot}/dx$, estimated to be $\approx $0.3 mT/nm [Fig.~\ref{fig:figS1}(b)], the electron is then subject to an oscillating magnetic field $B_{a.c.}=\frac{eE_{a.c.}l^2_{orb}|dB_{\bot}/dx|}{\Delta_{orb}}\propto\frac{eE_{a.c.}|dB_{\bot}/dx|}{\Delta^2_{orb}}$, perpendicular to the external static magnetic field. Here $\Delta_{orb}\propto 1/l^2_{orb}$ is the orbital level spacing, with $l_{orb}$ the typical QD dimension. We notice that the amplitude of $B_{a.c.}$ is proportional to the magnitude of the magnetic field gradient, to $1/\Delta_{orb}^2$ and to the amplitude of $E_{a.c.}$.

We tune the right dot to the few-electron regime (with the left dot empty) and adjust the tunnel rate between the dot and the reservoir to be around $\sim$1 kHz, so that dot-reservoir tunnel events can be monitored in real time by collecting the charge sensor current, read through a room temperature I-V converter, after a low-pass-filter with corner frequency of 30 kHz. The single-shot data \cite{Elzerman2004,Kawakami2014} are processed on the fly by using an FPGA which directly counts the number of spin excited state events by using a threshold detection scheme.

\subsection{Estimation of the ratio of Rabi frequencies from CW measurements}

\begin{figure}
\includegraphics[width=8.5cm] {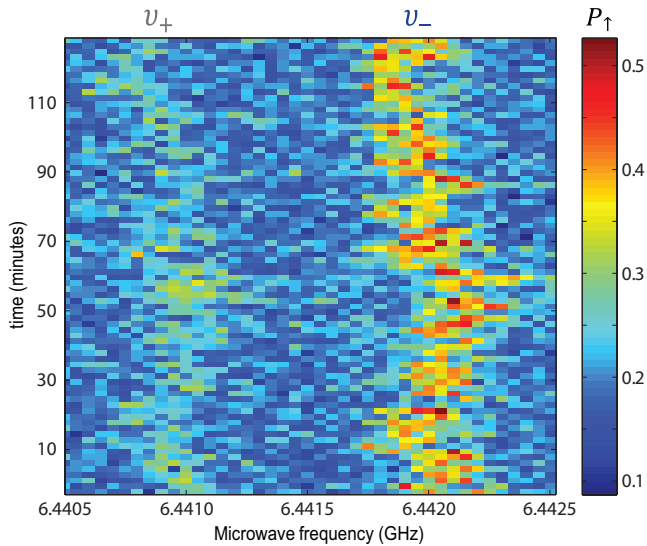}
\caption{\label{fig:figS2} The raw data on which Fig.~1(d) is based. Measured spin-up probability, $P_\uparrow$, as a function of applied microwave drive frequency $f_{MW}$ and time (external field $B_{ext}$= 560.783 mT, power P = -12 dBm, microwave pulse duration 700 $\mu$s). Each horizontal scan in the figure takes $\sim$2 minutes (200 cycles, in which each data point takes 2s), and the scan is repeated $\sim$60 times. The time reported on the y-axis is counted from the start of the measurement.}
\end{figure}

A typical CW spin resonance measurement is reported in Fig.~\ref{fig:figS2}, which shows the spin excited state probability as a function of the applied microwave frequency, $f_{MW}$, and time. We note that the fluctuations of the two spin resonance frequencies, $f^{(3)}$ and $f^{(4)}$, are highly correlated; this suggest that the two resonances are affected by the same low-frequency source of noise (very likely hyperfine fluctuations) on the $\sim$minute timescale. The trace for the second harmonic in Fig.~1(d) in the main text is obtained by averaging the data in Fig.~\ref{fig:figS2} over the time axis.

As reported in the supplementary information of Ref.~\cite{Kawakami2014}, the ratio of the steady-state spin-flip probability measured for the two valley states converges at low driving power to $r_R^2(2ph)$, with $r_R(2ph)$ the ratio of the Rabi frequencies between two valleys [$r_R(2ph)=f_1^{(4)}/f_1^{(3)}$].\\
The ratio of the measured peak amplitudes in Fig.~1(d), which we write as $A(\upsilon_+)/A(\upsilon_-)$, is the product of $r_R^2(2ph)$ and the ratio of the respective populations at the end of the initialization stage, $\epsilon^{(4)}/\epsilon^{(3)}$:
\begin{equation}
\frac{A(\upsilon_+)}{A(\upsilon_-)} = r_R^2(2ph)\times\frac{\epsilon^{(4)}}{\epsilon^{(3)}}
\label{eq:ratio peaks}
\end{equation}

From a Gaussian fit to the spin resonance response for the two valleys in Fig.~1(d), we extract $A(\upsilon_+)/A(\upsilon_-)\sim\,0.35$. 
Furthermore, it is reasonable to assume that the ratio between the valley ground and excited state populations after the initialization stage, is the same when driving via the second harmonic as when driving via the fundamental. Therefore,  we can use the ratio extracted  for the fundamental in \cite{Kawakami2014}:
\begin{equation}
\frac{\epsilon^{(4)}}{\epsilon^{(3)}}\equiv\frac{\epsilon^{(2)}}{\epsilon^{(1)}}=\frac{0.3\pm0.1}{0.7\pm0.1}=0.42\pm0.20.
\label{eq:rabi_ratio}
\end{equation}
From eq.~\ref{eq:ratio peaks} and eq.~\ref{eq:rabi_ratio} we obtain
\begin{equation}
r_R(2ph)=f_1^{(4)}/f_1^{(3)}=0.90\pm0.21.
\label{eq:rabi_ratio2}
\end{equation}
We remark that we also assume that the ratio of the Rabi frequencies between two valleys is the same at high microwave power as it is at low microwave power, which seems reasonable.

\subsection{Chevron pattern using second harmonic driving}

\begin{figure*}[ht!]
\includegraphics[width=6in]{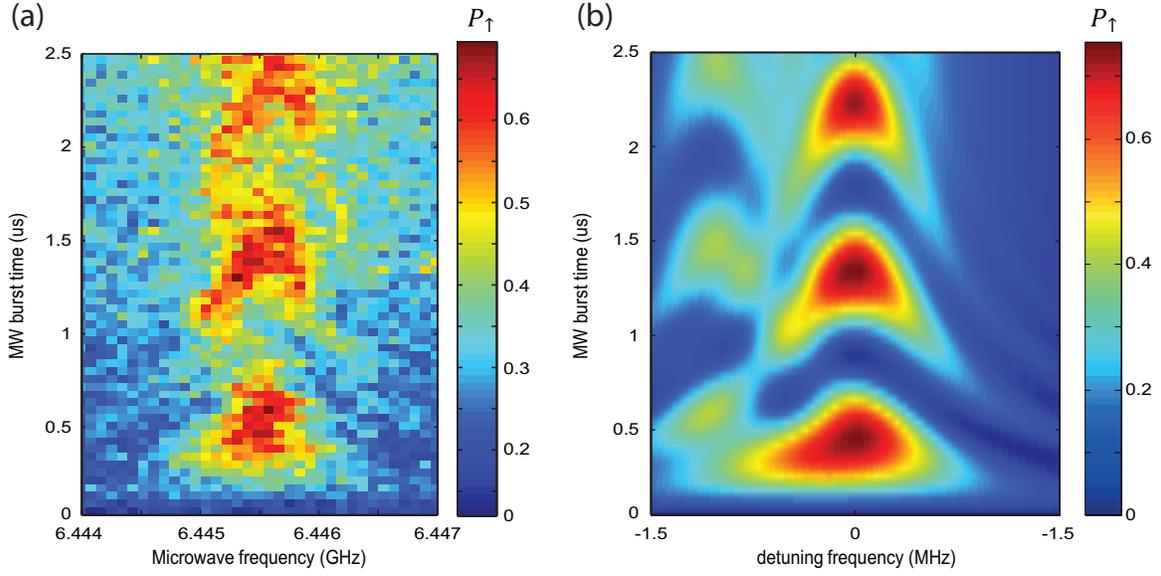}
\caption{(a) Measured spin-up probability, $P_\uparrow$, as a function of drive frequency $f_{MW}$ and microwave burst time ($B_{ext}$ = 560.783 mT, P = 11.0 dBm). (b) Simulated spin-up probability using population fractions 0.3:0.7, Rabi frequencies 1.0 MHz and 1.1 MHz for the $\upsilon_+$ and $\upsilon_-$ valley states respectively, and the Larmor frequencies separated by $\sim$1.04 MHz, extracted from low-power CW measurements in Fig.~1(d).
}
\label{fig:figRabi_det} 
\end{figure*}

In Fig.~\ref{fig:figRabi_det}(a) we report the spin excited state probability oscillations as a function of the microwave burst time and detuning frequency (driving with $P_{MW}$= 11.0 dBm at the source). The quality of the data (stability of the measurement) is not high enough to extract independently the Rabi frequencies for the two valley states directly from the superimposed Chevron patterns (as was done in \cite{Kawakami2014} for driving via the fundamental harmonic). However, using the Rabi frequency ratio $r_R(2ph)\approx 1$ extracted above, the ratio of initial populations of the two valleys of $\epsilon^{(4)}/ \epsilon^{(3)}\sim$0.3/0.7 discussed above as well, and a Rabi frequency of 1.05 MHz for the valley ground state [estimated from a FFT along the MW burst time of Fig.~\ref{fig:figRabi_det}(a)], we can simulate the two superimposed Chevron patterns, see Fig.~\ref{fig:figRabi_det}(b). This simulation agrees qualitatively with the data of Fig.~\ref{fig:figRabi_det}(a), in particular it captures the slight asymmetry along the detuning axis, and the fact that mostly one Chevron pattern is visible.

\subsection{T$_2^\ast$ estimation from Ramsey envelope decay and Ramsey simulation}

\begin{figure}
\includegraphics[width=8.5cm] {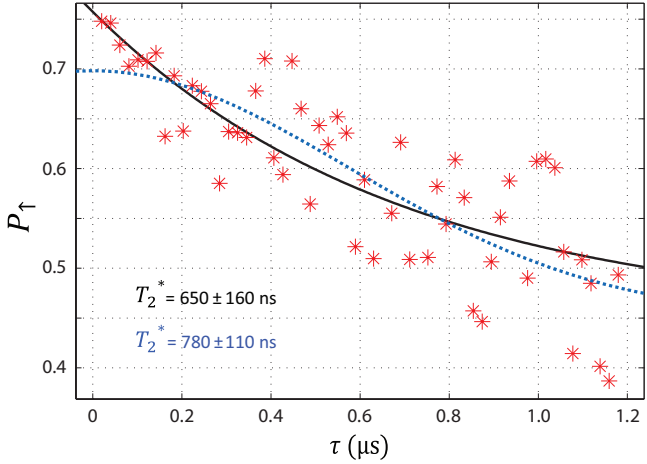}
\caption{\label{fig:Ramsey_decay} Decay of Ramsey envelope as a function of the waiting time between the two Ramsey pulses. The blue dotted line is a fit of the data with the relation $P_\uparrow=a \exp[-(t/T_2^\ast)^2]+c$, with $T_2^\ast$ as a free parameter and $c=0.46$, the average value for $\tau \gg T_2^\ast$ (see Fig.~\ref{fig:fig4}), which we also obtain keeping $c$ a free parameter ($c=$0.46$\pm$0.09; $T_2^\ast$= 790$\pm$330 ns). The fact that the center of the $P_\uparrow$ oscillations is not 0.5 is attributed to initialization and measurement errors.}
\end{figure}

In order to get an alternative estimation of the dephasing time $T_2^\ast$, we can perform a Ramsey measurement [see Fig.~\ref{fig:fig4}(a) of the main text] and record the amplitude of the Ramsey oscillations as a function of the waiting time $\tau$ between the two $\pi/2$ pulses, keeping $f_{MW}\approx f_0^{(1)}$. We show this data in Fig.~\ref{fig:Ramsey_decay}, with the blue dotted curve representing the fitting relation $P_\uparrow=a \exp[-(t/T_2^\ast)^2]+c$. From this fit we get a $T_2^\ast$ of 780 $\pm$ 110 ns, compatible with what we estimated from the CW spin resonance linewidth in Fig.~1(d). On the same figure we report in black for comparison a fit with the exponential relation $P_\uparrow=a \exp(-t/T_2^\ast)+c$.

\subsection{Echo decay of a qubit driven fundamental harmonic and comparison to results for driving at second harmonic}

\begin{figure}
\includegraphics[width=8.5cm] {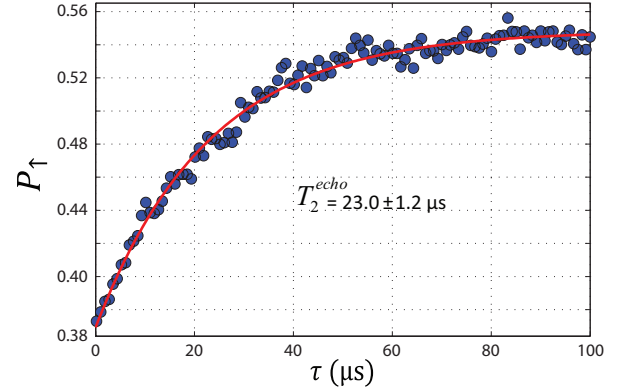}
\caption{\label{fig:figS5}  Spin echo measurement for the ground valley-orbit state driven at the fundamental harmonic for $B_{ext}$= 560.783 mT and $f_{MW}=f_0^{(1)}=2f_0^{(3)}$. The observed $T_2^{echo}$ times are similar for driving at the fundamental and second harmonic.}
\end{figure}

In Fig.~\ref{fig:figS5} we show a Hahn echo measurement realized driving spin resonance at the fundamental harmonic, recorded at the same magnetic field $B_{ext}$ as the measurement reported in Fig.~4(e) (driven by second harmonic). The $T^{echo}_2(1ph)$ extracted from a fit with a single exponential decay is similar to the $T_2^{echo}(2ph)$ extracted from Fig.~4(e). The decay obeys a single exponential similar to that observed in Ref.~\cite{Kawakami2014}, indicating the presence of a high-frequency decoherence process. 

\section{II. Mechanisms mediating second harmonic generation.} 

Here we elaborate on the three mechanisms discussed in the main text that can lead to second harmonic generation.

\subsection{Position dependent magnetic field gradient}

From a simulation of the stray magnetic field of the micromagnet, see also the supplementary information of Ref.~\cite{Kawakami2014}, we can put an upper bound on the second derivative of the stray magnetic field with respect to the dot coordinates of $d^2 B_\bot/dx^2 < 1$ $\mu$T/(nm)$^2$. This is far too small to produce second harmonic Rabi frequencies that are comparable to the fundamental harmonic Rabi frequencies (which is what we observed experimentally).

\subsection{Tilted drive field}

If the a.c. magnetic field that is driving spin resonance is not strictly perpendicular to the static magnetic field, second harmonic driving becomes possible~\cite{Gromov2000}. Eq. 18 of Ref.~\cite{Gromov2000} expresses the two-photon Rabi frequency, $\omega^{(2ph)}_{eff}$, as a function of the drive strength $\omega_1\propto B_{a.c.}$ and the drive frequency $\omega\propto B_{tot}$, for the case of an angle $\vartheta$ between static and oscillating field of 45 degrees. They found $\omega^{(2ph)}_{eff}=2\omega_1^2/\omega$. In our system the magnetic field gradient (which will provide $B_{a.c.}(t)$ in combination with the microwave electric field) has components in all three directions of space. More specifically, from \cite{Kawakami2014} we have $dB_{//}/dx \approx 0.2$ mT/nm and $dB_\bot/dx\approx 0.3$ mT/nm, from which we can estimate that $\vartheta\sim$ 56 degrees. Furthermore, from \cite{Kawakami2014} we also get $\omega_1\approx$ 3 MHz and $\omega\approx$ 13 GHz from which $\omega^{(2ph)}_{eff}/\omega^{(1ph)}=2\omega_1/\omega<<10^{-3}$, too small to explain the measured $f_1^{(3)}/f_1^{(1)}\sim 0.5$ for the highest microwave driving field used in the experiment [see Fig.~2(b)].

\subsection{Anharmonic confining potential}

Another possible mechanism which can give rice to the two-photon process in the Si/SiGe micromagnet-EDSR experiments can be related to the presence of an intermediate level (valley excited state spin down) lying between the initial and final states (valley ground spin down and up) which can mediate the resonance process [private communication A. P\'{a}lyi and M. Rudner]. If this is the case, by measuring the B-field dependence of the two-photon Rabi frequency, one should observe a non-monotonic $f_{Rabi}^{2ph}(B_{ext})$ dependence: the Rabi frequency should grow as half of the Larmor frequency approaches the valley splitting either from above or from below. Even if we did not perform a systematic study of the $f_{Rabi}^{2ph}$ vs $B_{ext}$, from the measurement we recorded we can exclude this striking dependence.

In the absence of detailed knowledge of the shape of the confining potential, it is difficult to quantitatively estimate the magnitude of this effect. However, what we can say is that the details of the confining potential strongly influence the efficiency of second harmonic driving: After tuning the device into a new gate voltage configuration (for which the valley splitting is higher than it is here), it has not been possible to observe again clear signatures of second harmonic driving. The new settings will likely also shift the average dot position in the micro magnet stray field, thereby altering the tilt of the effective driving field as well as the position dependent field gradient. However, as discussed in the preceding sections, these two effects are several orders of magnitude too small to explain the observed second harmonic driving. If the degree of anharmonicity of the confining potential dominates second harmonic driving, control of the anharmonicity will be required to make routine use of second harmonic driving in future spin qubit experiments.

\end{document}